\documentclass[11pt,a4paper]{article}

\usepackage[utf8]{inputenc}
\usepackage[T1]{fontenc}
\usepackage{textcomp}
\usepackage{amsmath,amssymb}
\usepackage{graphicx}
\usepackage{booktabs}
\usepackage{hyperref}
\usepackage[margin=1in]{geometry}
\usepackage{enumitem}
\usepackage{xcolor}
\usepackage{fancyhdr}
\usepackage{titlesec}
\usepackage{algorithm}
\usepackage{algpseudocode}
\usepackage{multirow}

\title{\textbf{3D Cloud Component Analysis}\\[4pt]
\textbf{of Atomic Structures}}
\author{Pai Li\\[4pt]
\small State Key Laboratory of Materials for Integrated Circuits,\\
\small Shanghai Institute of Microsystem and Information Technology,\\
\small Chinese Academy of Sciences, 865 Changning Road, Shanghai 200050, China\\[4pt]
\small \texttt{lipai@mail.sim.ac.cn}}
\date{August 2026}

\begin{document}
\maketitle

\begin{abstract}
We present a method for decomposing atomic structures into physically meaningful components by converting discrete atomic coordinates into continuous three-dimensional density fields. Each element is represented by a Gaussian-smeared density map with values ranging from 0 to 1, computed efficiently through a bin-then-blur approach with periodic boundary conditions. The sum of all element densities, truncated at unity, defines the material region; its complement defines the vacuum. Every voxel is first assigned a chemical formula from the set of elements present above a threshold; the resulting formula map is then cleaned so that only regions with a genuine bulk interior---measured by the Euclidean distance to their own boundary---survive as components. Thin surface terminations (e.g., a Ga monolayer on GaAs) and one-to-two-voxel boundary layers between two crystals are absorbed by the neighboring stable region, so the decomposition contains exactly the bulk-like chemical-formula components and the vacuum, with no surface or interface components. For each component we determine the crystal phase with a neural prototype classifier: one phase if the trusted interior atoms vote unanimously, two phases (e.g., crystalline and amorphous Si) if they split into two confident groups, in which case the component is divided into two. Interfaces and surfaces are then derived as boundaries---material--material and material--vacuum---and every atom is labeled bulk, surface, interface, or vertex, with vertex reserved for geometric corners of a component. Inside crystalline components, inner defects are detected from coordination-number and local-density deviations. We demonstrate the approach on Si/GaAs and Si/SiO$_2$ heterojunctions, crystalline/amorphous silicon junctions, vacancy-containing crystals, and bulk crystals.
\end{abstract}

\section{Introduction}

\subsection{Motivation}

Atomic structures in computational materials science are conventionally represented as discrete point sets: each atom is a coordinate triple associated with an element symbol and a lattice vector. This representation is exact for input to density functional theory (DFT) and molecular dynamics codes, but it is poorly suited for answering compositional questions that humans naturally ask about a structure: \textit{Where does the silicon end and the silicon dioxide begin? What is the chemical composition of the interface region? Is there a continuous path of vacuum through this porous material? Which atoms are on the surface, and which are buried in the bulk?}

Answering such questions requires a representation that makes spatial extent and chemical mixing explicit. A discrete atom list does not directly encode the \textit{region} that an atom occupies---its sphere of influence, the volume it contributes to a phase, or the degree to which it mixes with neighboring atoms of different elements.

The need for such compositional analysis is sharpened by the rise of language-model-driven materials workflows, which generate structures from natural-language descriptions via multi-agent pipelines~\cite{multiagent2025,llm_crystals,genms2024}. These pipelines produce structures at scale---heterojunctions, coated particles, doped supercells---that must be checked against the requested composition before any downstream simulation, yet the check is exactly the step a discrete atom list makes tedious.

We address this by converting discrete atomic coordinates into continuous three-dimensional density fields via Gaussian smearing. Each atom is replaced by a normalized Gaussian centered at its position, producing a smooth density function $\rho_e(\mathbf{r})$ for each element $e$. The sum of element densities (clipped to $[0,1]$) defines where material exists; its complement defines vacuum. Overlapping element densities naturally reveal chemical mixing at interfaces, grain boundaries, and disordered regions. Connected-component analysis of this field partitions the structure into physically meaningful regions.

\subsection{Relation to Prior Work}

Our approach draws on two traditions. The first is \textbf{Gaussian density representations} used in machine learning interatomic potentials~\cite{behler2011,shapeev2016}, where atomic environments are encoded as sums of Gaussian or polynomial basis functions for symmetry-function construction. We adopt the Gaussian smearing machinery but repurpose it for spatial segmentation rather than energy prediction.

The second is \textbf{voxel-based materials analysis}, exemplified by tools such as Zeo++~\cite{zeopp} for pore characterization and PoreBlazer~\cite{poreblazer} for porosity analysis. These tools operate on void spaces in zeolites and metal-organic frameworks. Our method is more general: it simultaneously characterizes both material and vacuum regions, handles multi-element systems with compound interface detection, and provides per-atom spatial classification.

Existing general-purpose tools occupy different niches: the Atomic Simulation Environment~\cite{ASE} provides programmatic structure manipulation but no spatial segmentation; OVITO~\cite{OVITO} and VESTA~\cite{VESTA} excel at visualization and manual inspection, with per-atom classification limited to coordination and bond-order quantities; Atomsk~\cite{atomsk} is a structure builder and converter. None of these answers the compositional questions above---where a phase begins and ends, what the interface chemistry is, whether a void percolates---in a machine-consumable form.

The method is implemented in a modular Python package (\texttt{cloud\_comp}) with a Dash-based web interface following the architecture of the ATLAS framework~\cite{atlas}, supporting CIF, POSCAR/VASP, and XYZ input formats via pymatgen~\cite{pymatgen}.

\section{Algorithm}

\subsection{Pipeline Overview}

The analysis proceeds in seven stages:

\begin{enumerate}[label=\textbf{S\arabic*}. ,leftmargin=*]
    \item \textbf{Structure Loading}: Parse atomic structure (CIF, POSCAR/VASP, XYZ) via pymatgen, extracting lattice matrix $\mathbf{L} \in \mathbb{R}^{3\times 3}$, fractional coordinates $\mathbf{f}_i$, and element types.
    \item \textbf{3D Density Grid}: For each element, bin atoms onto a uniform $n_z \times n_y \times n_x$ grid spanning one unit cell, then apply 3D Gaussian blur with periodic boundary conditions. Preserve both raw (unclipped) and clipped ($[0,1]$) density maps.
    \item \textbf{Chemical-Formula Map}: Classify each grid voxel as vacuum, pure element, or compound from the set of elements above the density threshold. This is a formula map, not a component map: a Ga-terminated GaAs surface and the GaAs bulk share the same crystal but appear as different voxel classes.
    \item \textbf{Formula-Region Cleaning}: Measure the Euclidean depth of every connected class region; regions without a bulk interior (surface monolayers, one-to-two-voxel boundary layers, noise) are absorbed into the neighboring stable region. The cleaned map contains only bulk-like chemical-formula regions plus vacuum---there are no surface or interface components.
    \item \textbf{Connected Component Labeling}: Partition the cleaned formula map into connected components with periodic boundary awareness via union-find merging.
    \item \textbf{Phase Analysis}: For each component, classify its trusted interior atoms with a neural prototype classifier. A component with a single confident prototype keeps that phase; a component whose trusted votes split into two confident groups (e.g., crystalline and amorphous Si) is divided into two components, one per phase.
    \item \textbf{Characterization}: Compute component properties (volume, stoichiometry, shape descriptors, cavity/tunnel counts), derive interfaces and surfaces as boundaries between components, classify every atom as bulk/surface/interface/vertex, index surface facets and interfaces with Miller indices recovered from the component lattice (Section~\ref{sec:lattice}), and flag inner defects from coordination-number and local-density deviations.
\end{enumerate}

Surfaces and interfaces are therefore \emph{relations between components}, never components themselves: a surface is the boundary between a material component and the vacuum component, and an interface is the boundary between two material components. This separation keeps the component list physically interpretable---each entry is a region of one chemical formula and one crystal phase---while the derived boundary information supplies the per-atom labels needed for device and interface analysis.

\subsection{3D Gaussian Density Grid}

\subsubsection{Grid Construction}

A uniform Cartesian grid is constructed over the unit cell. Given a target spacing $\Delta$ (default 0.25~\AA), the grid dimensions are:

\begin{equation}
n_x = \left\lceil \frac{|\mathbf{a}|}{\Delta} \right\rceil, \quad
n_y = \left\lceil \frac{|\mathbf{b}|}{\Delta} \right\rceil, \quad
n_z = \left\lceil \frac{|\mathbf{c}|}{\Delta} \right\rceil
\end{equation}

capped at $n_{\max}=300$ per dimension (~200~MB total memory for all density arrays). Grid centers are at:

\begin{equation}
x_i = \left(i + \tfrac{1}{2}\right) \frac{|\mathbf{a}|}{n_x}, \quad
y_j = \left(j + \tfrac{1}{2}\right) \frac{|\mathbf{b}|}{n_y}, \quad
z_k = \left(k + \tfrac{1}{2}\right) \frac{|\mathbf{c}|}{n_z}
\end{equation}

Orthogonal cells are handled directly with \texttt{mode=`wrap'} in the Gaussian filter. Non-orthogonal (triclinic) cells are handled by a fallback path: atoms are replicated into neighboring cells in an expanded grid, the Gaussian filter is applied with \texttt{mode=`constant'}, and the central unit cell is extracted. This costs a constant factor in memory and time but supports arbitrary lattice geometries.

\subsubsection{Bin-then-Blur Algorithm}

The naive approach---evaluating $\exp(-|\mathbf{r} - \mathbf{r}_i|^2 / 2\sigma^2)$ at every grid point for every atom---has complexity $O(N_{\text{atoms}} \times n_x n_y n_z)$, which is prohibitive for large structures. We instead use a two-stage ``bin-then-blur'' approach with complexity $O(N_{\text{atoms}} + n_x n_y n_z)$:

\begin{algorithm}[ht]
\caption{Compute Element Density Grid}
\label{alg:density}
\begin{algorithmic}[1]
\Require Atoms of element $e$ at Cartesian positions $\{\mathbf{r}_i\}_{i=1}^{N_e}$, grid dimensions $(n_z, n_y, n_x)$, Gaussian width $\sigma$, cell extents $(L_x, L_y, L_z)$
\Ensure Density grid $\boldsymbol{\rho}_e \in [0,1]^{n_z \times n_y \times n_x}$ (clipped), $\boldsymbol{\rho}_e^{\text{raw}} \in \mathbb{R}_{\geq 0}^{n_z \times n_y \times n_x}$ (unclipped)
\State $\mathbf{G} \gets \mathbf{0}_{n_z \times n_y \times n_x}$
\For{each atom position $\mathbf{r}_i = (x_i, y_i, z_i)$}
    \State $ix \gets \lfloor x_i / L_x \cdot n_x \rfloor \bmod n_x$
    \State $iy \gets \lfloor y_i / L_y \cdot n_y \rfloor \bmod n_y$
    \State $iz \gets \lfloor z_i / L_z \cdot n_z \rfloor \bmod n_z$
    \State $\mathbf{G}[iz, iy, ix] \gets \mathbf{G}[iz, iy, ix] + 1.0$ \Comment{Bin}
\EndFor
\State $\boldsymbol{\rho}_e^{\text{raw}} \gets \text{GaussianFilter3D}(\mathbf{G}, \boldsymbol{\sigma}=(\sigma/\Delta_z, \sigma/\Delta_y, \sigma/\Delta_x), \text{mode}=\text{`wrap'})$ \Comment{Blur}
\State $\boldsymbol{\rho}_e \gets \text{clip}(\boldsymbol{\rho}_e^{\text{raw}}, 0, 1)$ \Comment{Clip}
\end{algorithmic}
\end{algorithm}

The \texttt{GaussianFilter3D} is implemented via \texttt{scipy.ndimage.gaussian\_filter} with \texttt{mode=`wrap'}. The wrap mode causes the Gaussian kernel to sample from the opposite side of the grid when it extends beyond a boundary, which is physically exact for periodic systems: density from an atom near one cell face correctly spills across to the opposite face. The pixel-unit sigmas are $\sigma/\Delta_x$, $\sigma/\Delta_y$, $\sigma/\Delta_z$.

\subsubsection{Dual Representation}

We maintain two copies of each element's density map:

\begin{itemize}
    \item \textbf{Raw} $\boldsymbol{\rho}_e^{\text{raw}}$: The unclipped output of the Gaussian filter. Preserves the true integral of Gaussian contributions, which is essential for stoichiometry computation (Section~\ref{sec:stoichiometry}).
    \item \textbf{Clipped} $\boldsymbol{\rho}_e = \min(1, \boldsymbol{\rho}_e^{\text{raw}})$: Values constrained to $[0,1]$. Used for all shape-related computations: vacuum detection, voxel classification, and component boundary definition.
\end{itemize}

The total material density and vacuum density are computed from clipped maps:

\begin{align}
\boldsymbol{\rho}_{\text{total}}[i,j,k] &= \min\!\left(1,\; \sum_e \boldsymbol{\rho}_e[i,j,k]\right) \label{eq:total}\\
\boldsymbol{\rho}_{\text{vacuum}}[i,j,k] &= 1 - \boldsymbol{\rho}_{\text{total}}[i,j,k] \label{eq:vacuum}
\end{align}

By construction, $\boldsymbol{\rho}_{\text{total}} + \boldsymbol{\rho}_{\text{vacuum}} = 1$ everywhere, and vacuum dominates precisely where material is absent.

\subsection{Chemical-Formula Map}

\label{sec:classification}

Each voxel is assigned a chemical formula from the set of elements present in its density profile:

\begin{equation}
\text{class}[i,j,k] =
\begin{cases}
0 \text{ (vacuum)} & \text{if } \boldsymbol{\rho}_{\text{vacuum}}[i,j,k] > \theta_{\text{vac}} \\
\text{element } e & \text{if } |\{e : \boldsymbol{\rho}_e[i,j,k] > \theta_{\text{elem}}\}| = 1 \text{ and } \{e\} = \{X\} \\
\text{compound } \{e_1, \ldots, e_m\} & \text{if } |\{e : \boldsymbol{\rho}_e[i,j,k] > \theta_{\text{elem}}\}| \geq 2
\end{cases}
\end{equation}

where $\theta_{\text{vac}} = 0.5$ is the vacuum threshold and $\theta_{\text{elem}} = 0.3$ is the element presence threshold. A voxel with no element above $\theta_{\text{elem}}$ but vacuum density below $\theta_{\text{vac}}$ is an edge case classified as vacuum.

Class IDs are assigned as follows: 0 for vacuum, 1--99 for pure elements (dynamically mapped per structure), and 100+ for compounds (hashed from the sorted tuple of constituent element symbols). The result is a voxel-level \emph{formula map}: two adjacent voxels carry the same label only if the same set of elements is present.

The element threshold $\theta_{\text{elem}} = 0.3$ is chosen to balance two competing requirements: (i) it must be high enough to avoid classifying random noise tails as compound regions, and (ii) it must be low enough to detect genuine interface overlap where both elements contribute at roughly 30\% of their peak density. For a Gaussian of width $\sigma = 1.5$~\AA, two atoms at a typical bond distance of $\sim$2~\AA\ each contribute $\exp(-1^2 / (2 \cdot 1.5^2)) \approx 0.8$ at the midpoint---both well above 0.3---so the interface is correctly identified as a compound.

\subsubsection{Why the raw formula map is not the component map}

The threshold classification is intentionally local and does not yet know what is a material and what is merely a boundary. Three artifacts follow for compound crystals:

\begin{itemize}
    \item \textbf{Surface terminations}: on a zincblende (100) or (110) surface the outermost atomic layer is a single element. In the density field this layer is a thin shell whose formula differs from the bulk (pure Ga or pure As on GaAs), even though it belongs to the same crystal.
    \item \textbf{Boundary layers}: between two materials the two density fields overlap over one to two voxels, producing mixed-formula shells (e.g., GaSi or GaAsSi between Si and GaAs).
    \item \textbf{Noise}: isolated one-voxel regions appear wherever two density tails cross the threshold independently.
\end{itemize}

If these shells were labeled as components, the result would contain spurious ``pure Ga'', ``pure As'' and ``interface'' components, and atoms at genuine free surfaces would appear to touch a second material. We remove the artifacts at the voxel level instead.

\subsubsection{Stable regions}

\label{sec:stableregions}

A connected region of one formula is \emph{stable} if it has a genuine bulk interior. We measure the interior with the Euclidean distance transform of the region's own mask, $\text{EDT}(\mathbf{M})$: the value at each voxel is the distance (in \AA) to the nearest voxel not in the region. The region's interior depth is

\begin{equation}
h(\mathcal{R}) = \max_{(i,j,k) \in \mathcal{R}} \text{EDT}(\mathbf{M}_{\mathcal{R}})[i,j,k],
\end{equation}

and the region is stable if $h(\mathcal{R}) \geq \tau_{\text{stab}}$ with $\tau_{\text{stab}} = 2.0$~\AA\ by default. For $\sigma = 1.5$~\AA\ and $\Delta = 0.3$~\AA\ this cleanly separates ~10~\AA\ thick crystals ($h \approx 10$~\AA) from surface and boundary shells ($h \approx 1.0$--1.5~\AA). A genuine thin film thicker than $\tau_{\text{stab}}$ retains its own formula and becomes its own component.

\subsubsection{Absorption of unstable regions}

Every unstable region is absorbed into the neighboring stable region with which it shares the largest face-contact area. Because an absorbed region inherits the stable region's formula, chains of shells (e.g., a GaSi layer touching both a Si and a GaAsSi shell) are resolved iteratively until no unstable region remains. The assignment is purely geometric and uses only the formula map; no classifier output is involved.

After cleaning, the formula map contains exactly the bulk-like chemical-formula regions plus vacuum. A surface monolayer on GaAs is absorbed by GaAs; a one-voxel boundary layer between Si and GaAs is absorbed by whichever side it touches most. There are consequently \emph{no surface components and no interface components} in the final decomposition: surfaces and interfaces are boundaries between the surviving components (Section~\ref{sec:boundaries}).

\subsection{Connected Component Labeling with PBC}

\label{sec:components}

Standard 3D connected-component algorithms (e.g., \texttt{scipy.ndimage.label}) do not account for periodic boundary conditions. A component that spans a cell boundary---such as a bulk crystal that wraps around in all three directions---would be incorrectly split into separate pieces.

We solve this through a two-stage approach:

\textbf{Stage 1: Standard Labeling.} Run \texttt{scipy.ndimage.label} with a 6-connectivity (face-sharing) structure on the classification array. All non-zero (material) voxels are treated as foreground. This produces $N_{\text{temp}}$ provisional labels. The 6-connectivity is preferred over 26-connectivity because corner-touching voxels in a density-based segmentation typically correspond to physically distinct regions connected only by a thin bridge; face-sharing connectivity produces more physically meaningful components.

\textbf{Stage 2: PBC Merging via Union-Find.} For each periodic axis, examine opposing face pairs:

\begin{itemize}
    \item $z$-faces: compare $\mathbf{L}[0, :, :]$ with $\mathbf{L}[n_z-1, :, :]$
    \item $y$-faces: compare $\mathbf{L}[:, 0, :]$ with $\mathbf{L}[:, n_y-1, :]$
    \item $x$-faces: compare $\mathbf{L}[:, :, 0]$ with $\mathbf{L}[:, :, n_x-1]$
\end{itemize}

For each pair of opposing voxels $(i,j,k)$ and $(i',j',k')$ on opposite faces, if both are foreground and have different provisional labels $l_a \neq l_b$, we record a union operation $\text{UF}.\text{union}(l_a, l_b)$. Edge pairs (intersections of two periodic faces, e.g., the lines at $x=0,z=0$ vs.\ $x=n_x-1,z=n_z-1$) are also checked; corner pairs are handled transitively by the union-find data structure.

After processing all periodic faces, each provisional label is replaced by its union-find root, producing the final component labels. This approach avoids the $27\times$ memory cost of tiling a $3\times3\times3$ supercell.

\subsection{Phase Analysis}

\label{sec:phase}

The formula decomposition identifies \emph{what} elements are present, but not \emph{which crystal phase} they form: diamond Si and amorphous Si produce nearly identical density fields, and so do quartz and amorphous SiO$_2$. Phase identity is supplied by a neural prototype classifier trained on element-mapped, rotation-invariant local descriptors (Section~\ref{sec:implementation}) with 271 AFLOW prototypes plus an amorphous class. The classifier is invoked once per structure and returns, for every atom, a prototype label, a softmax confidence, an amorphous probability, and a per-atom nearest-neighbor distance $d_0$ that defines the descriptor window $R = 3\,d_0$ ($\approx 7$~\AA\ for Si and GaAs).

\subsubsection{Trusted interior atoms}

The classifier was trained on complete bulk environments. An atom whose window is truncated by a surface or interface---or corrupted by a periodic-seam artifact---produces a confident but wrong prototype (surface GaAs atoms are typically predicted ``amorphous'' or a random hexagonal prototype). Each atom is therefore tested for a \emph{complete local environment}:

\begin{enumerate}
    \item its full window $3\,d_0$ lies inside its component: the Euclidean distance from the atom's voxel to the component boundary is at least $3\,d_0$ (computed with the same distance transform used for region stability, Section~\ref{sec:stableregions});
    \item no short contacts: neither center--neighbor nor neighbor--neighbor distances fall below $0.6\,d_0$ (this catches wrapped slabs whose faces interpenetrate at the periodic seam).
\end{enumerate}

Atoms passing both tests are \emph{trusted}; only their predictions are allowed to vote.

\subsubsection{One phase or two}

For each component we count the prototype votes of its trusted atoms. Three outcomes are possible:

\begin{itemize}
    \item \textbf{One phase}: a single prototype dominates. The component keeps that prototype (e.g., \texttt{A4\_C\_diamond} for crystalline Si, \texttt{B3\_ZnS\_zincblende} for GaAs, \texttt{amorphous} for disordered material). Amorphous is a first-class phase: a formula component whose trusted interior votes are confidently amorphous is labeled amorphous, not crystalline.
    \item \textbf{Two phases}: the trusted votes split into two groups, each supported by at least $\phi_{\text{split}} = 20\%$ of the votes, at least $N_{\text{split}} = 10$ atoms, and a mean top-class confidence of at least $c_{\text{split}} = 0.6$. This is exactly the situation of coexisting crystalline and amorphous Si. The component is divided into two components of the same formula, one per phase; every atom joins the group of its nearest trusted atom of either phase. The confidence gate prevents scattered low-confidence votes in genuinely amorphous material from creating spurious prototype components.
    \item \textbf{No phase claim}: a component without trusted atoms (a slab thinner than the classifier window, a small cluster) has no reliable vote. Its phase is reported as \texttt{unknown} rather than a noisy majority.
\end{itemize}

The final prototype of \emph{every} atom of a component---inner, surface, interface, or vertex---is the component's phase, so classifier noise on individual boundary atoms is never used as a final label. Per-atom raw predictions and confidence are retained as metadata (\texttt{phase\_trusted}, \texttt{phase\_confidence}).

\subsection{Component Properties}

\label{sec:properties}

\subsubsection{Stoichiometry from Raw Density}
\label{sec:stoichiometry}

Clipping element densities to $[0,1]$ destroys the information needed to determine chemical formulas. In a SiO$_2$ region, both Si and O densities may saturate at 1.0, suggesting an incorrect 1:1 ratio. The solution is to integrate the \textit{raw} (unclipped) density maps over the component:

\begin{equation}
A_e = \sum_{(i,j,k) \in \mathcal{C}} \boldsymbol{\rho}_e^{\text{raw}}[i,j,k]
\end{equation}

where $\mathcal{C}$ is the set of voxel indices belonging to the component. Since the Gaussian width $\sigma$ is identical for all elements, each atom contributes exactly the same total integral $\int \exp(-r^2/2\sigma^2) \, d^3\mathbf{r} = (2\pi\sigma^2)^{3/2}$ to its element's raw density map. Therefore the ratio $A_{e_1} : A_{e_2} : \cdots$ directly gives the atom count ratio:

\begin{equation}
\frac{N_{e_1}}{N_{e_2}} = \frac{A_{e_1}}{A_{e_2}}
\end{equation}

Atoms near the component boundary have their Gaussian tails truncated, but because $\sigma$ is element-independent, the truncation fraction is identical for all elements and the ratio is preserved.

As a cross-validation, we also count atoms directly by assigning each atom to the component whose region contains its center position (Section~\ref{sec:atomclass}).

The estimated formula string is derived by normalizing the stoichiometric fractions, scaling to make the smallest fraction equal to 1, rounding to the nearest integer, and simplifying by the greatest common divisor.

\subsubsection{Shape Classification}

\label{sec:shape}

Shape characterization proceeds at three levels of detail:

\textbf{Level 1: Dimensionality.} For each of the three lattice directions, we determine whether the component is ``periodic'' (connected across the cell boundary) using absolute rather than fractional thresholds to avoid the scale-dependence problem. A component is periodic in direction $d$ if:

\begin{enumerate}
    \item \textbf{Gap check}: The largest gap between sorted voxel positions along direction $d$ (including the wrap-around gap) does not exceed $\delta_{\text{gap}} = 4.0$~\AA. This threshold corresponds to approximately twice a typical covalent bond length---any gap larger than this is physically meaningful.
    \item \textbf{Boundary check}: The component touches both cell boundaries in direction $d$ (minimum coordinate $\leq \delta_{\text{boundary}} = 2.0$~\AA\ and maximum coordinate $\geq L_d - \delta_{\text{boundary}}$).
\end{enumerate}

Using absolute thresholds avoids the pitfall of fractional thresholds, where a 20~\AA\ vacuum gap in a 100~\AA\ cell (20\% of the cell) would be missed by a 30\% threshold. The dimensionality is the count of periodic directions:

\begin{center}
\begin{tabular}{ccl}
\toprule
Periodic Directions & Shape Type & Example \\
\midrule
3 & bulk & Embedded crystal, dense polycrystal \\
2 & slab & Surface layer, heterojunction stack \\
1 & wire & Nanowire, nanotube, GAA channel \\
0 & nanoparticle & Isolated cluster, fullerene, quantum dot \\
\bottomrule
\end{tabular}
\end{center}

\textbf{Level 2: Shape Descriptors.} Five quantitative descriptors are computed:

\begin{itemize}
    \item \textbf{Aspect ratios}: $R_{xy} = L_x^{\text{bbox}} / L_y^{\text{bbox}}$, $R_{xz} = L_x^{\text{bbox}} / L_z^{\text{bbox}}$ from the bounding box extents.
    \item \textbf{Compactness}: $\gamma = V / V_{\text{bbox}}$, the fraction of the bounding box occupied by the component. $\gamma \to 1$ for a compact block; $\gamma \ll 1$ for a sparse or irregular shape.
    \item \textbf{Sphericity}: $\psi = \pi^{1/3} (6V)^{2/3} / A$, where $A$ is the surface area estimated from surface voxel count. $\psi = 1$ for a perfect sphere; lower values indicate deviation from sphericity.
    \item \textbf{Moment of inertia ratios}: The eigenvalues $\lambda_1 \geq \lambda_2 \geq \lambda_3$ of the inertia tensor $\mathbf{I} = \sum_i (\mathbf{r}_i - \bar{\mathbf{r}})(\mathbf{r}_i - \bar{\mathbf{r}})^\top$. The ratios $\lambda_2/\lambda_1$ and $\lambda_3/\lambda_1$ characterize elongation: both $\to 1$ for a sphere, $\lambda_2/\lambda_1 \to 1, \lambda_3/\lambda_1 \to 0$ for a disk, $\lambda_2/\lambda_1 \to 0, \lambda_3/\lambda_1 \to 0$ for a rod.
\end{itemize}

\textbf{Level 3: Topological Features.} We detect cavities (fully enclosed vacuum pockets) and tunnels (vacuum channels that percolate through the component):

\begin{algorithm}[ht]
\caption{Detect Cavities and Tunnels}
\label{alg:cavities}
\begin{algorithmic}[1]
\Require Component mask $\mathbf{M}$, full component labels $\mathbf{L}$, density grid
\Ensure Cavity count $n_{\text{cav}}$, tunnel count $n_{\text{tun}}$, cavity volumes
\State $\mathbf{V} \gets (\mathbf{L} = 0)$ \Comment{All vacuum voxels}
\State $\mathbf{V}_{\text{labels}}, N_V \gets \text{Label}(\mathbf{V})$ \Comment{Connected vacuum components}
\State $\mathcal{B} \gets \{\ell : \text{vacuum component } \ell \text{ touches any cell boundary}\}$
\State $n_{\text{cav}} \gets 0$
\For{$\ell = 1$ to $N_V$}
    \If{$\ell \notin \mathcal{B}$ \textbf{and} $\mathbf{V}_\ell$ adjacent to $\mathbf{M}$}
        \State $n_{\text{cav}} \gets n_{\text{cav}} + 1$ \Comment{Enclosed cavity}
    \EndIf
\EndFor
\For{$\ell \in \mathcal{B}$}
    \If{$\mathbf{V}_\ell$ adjacent to $\mathbf{M}$ \textbf{and} touches opposite cell faces}
        \State $n_{\text{tun}} \gets n_{\text{tun}} + 1$ \Comment{Tunnel}
    \EndIf
\EndFor
\end{algorithmic}
\end{algorithm}

A cavity is a connected vacuum component that does not touch any cell boundary but is adjacent to the material component (i.e., fully enclosed by it). A tunnel is a vacuum component that touches two opposite cell faces and passes through the component's bounding box extent.

\textbf{Level 4: Detailed Shape Label.} Based on dimensionality, cavity/tunnel counts, and geometric descriptors, a human-readable label is assigned:

\begin{center}
\small
\begin{tabular}{lll}
\toprule
Dimensionality & Features & Label \\
\midrule
\multirow{4}{*}{3 (bulk)}
    & No cavities, $\psi > 0.9$ & spherical\_bulk \\
    & $\geq 2$ tunnel directions & bicontinuous \\
    & 1 tunnel direction & tunneled\_bulk \\
    & $\geq 1$ cavity & porous\_bulk / hollow\_bulk / foam \\
    & $\lambda_3/\lambda_1 < 0.3$ & layered\_bulk \\
    & otherwise & compact\_bulk \\
\midrule
\multirow{3}{*}{2 (slab)}
    & Has tunnels & perforated\_slab \\
    & Aspect ratio $> 5$ & ribbon \\
    & $\psi < 0.3$ & flat\_plate \\
\midrule
\multirow{3}{*}{1 (wire)}
    & Has cavities & hollow\_wire (nanotube-like) \\
    & Aspect ratio $> 8$ & nanowire \\
    & otherwise & rod / column \\
\midrule
\multirow{5}{*}{0 (particle)}
    & Has cavities, $\psi > 0.8$ & hollow\_sphere \\
    & Has cavities & porous\_particle \\
    & $\psi > 0.9$ & sphere \\
    & $\psi > 0.7$ & spheroid \\
    & Aspect ratio $> 3$ & nanorod / disk \\
\midrule
\bottomrule
\end{tabular}
\end{center}

\subsubsection{Vacuum Shape}

For components classified as vacuum, we characterize their morphology in terms of connectivity to the cell boundary. Let $n_{\text{faces}}$ be the number of cell faces (out of 6) touched by the vacuum component, and $n_{\text{opp}}$ be the number of opposite-face pairs among them:

\begin{center}
\begin{tabular}{cccl}
\toprule
$n_{\text{faces}}$ & $n_{\text{opp}}$ & Label & Physical Meaning \\
\midrule
0 & 0 & pore & Enclosed void, isolated from boundaries \\
1 & 0 & surface\_vacuum & Vacuum bounding one side of a slab \\
2 & 0 & corner\_vacuum & Vacuum at a cell corner \\
2 & 1 & channel & Tunnel through the material in one direction \\
$\geq 3$ & $\geq 2$ & bicontinuous & Percolating in multiple directions \\
\bottomrule
\end{tabular}
\end{center}

\subsubsection{Per-Voxel Environment}
\label{sec:boundaries}

Every voxel in each component is classified by examining its 26 neighbors in the component label map:

\begin{itemize}
    \item \textbf{bulk}: All 26 neighbors belong to the same component.
    \item \textbf{surface}: At least one neighbor is vacuum (component 0).
    \item \textbf{interface}: At least one neighbor belongs to a different material component.
    \item \textbf{vertex}: A geometric corner: three or more of the six axis directions lead through the component's own density tail into open vacuum (Section~\ref{sec:atomclass}).
\end{itemize}

Interface takes priority over surface and vertex: a voxel that touches vacuum \emph{and} another material is an interface voxel. Vertex is a property of a single component's geometry (where three surface facets meet) and is never an interface voxel. Surfaces and interfaces are therefore boundaries of the component---material--vacuum and material--material---and the per-component counts of surface, interface and vertex voxels feed the surface-area and interface-area estimates.

\subsection{Atom Classification}
\label{sec:atomclass}

Each atom in the original structure is mapped to a component and classified by its local environment.

\subsubsection{Component Assignment}

For atom $i$ at fractional coordinates $(f_a, f_b, f_c)$, the grid index is:

\begin{equation}
ix = \lfloor f_a \cdot n_x \rfloor \bmod n_x, \quad
iy = \lfloor f_b \cdot n_y \rfloor \bmod n_y, \quad
iz = \lfloor f_c \cdot n_z \rfloor \bmod n_z
\end{equation}

The atom is assigned to $\texttt{component\_labels}[iz, iy, ix]$. This is an $O(1)$ lookup per atom.

\subsubsection{Local Environment Classification}

A spherical neighborhood of radius $r_{\text{class}} = 3.0$~\AA\ (configurable) is sampled around the atom in the component label map. Let $n_{\text{same}}$ be the count of voxels in the same component, $n_{\text{vac}}$ the count in vacuum, and $\mathcal{S}_{\text{other}}$ the set of distinct other material component IDs:

\begin{equation}
\text{classification} =
\begin{cases}
\text{bulk} & \text{if } n_{\text{vac}} = 0 \text{ and } |\mathcal{S}_{\text{other}}| = 0 \\
\text{interface} & \text{if } |\mathcal{S}_{\text{other}}| > 0 \quad \text{(priority)} \\
\text{surface} & \text{if } n_{\text{vac}} > 0 \\
\text{vertex} & \text{otherwise, if } \geq 3 \text{ free surface directions}
\end{cases}
\end{equation}

Interface takes priority over surface and vertex, matching the per-voxel rule: an atom that reaches both vacuum and a second material is an interface atom. \textbf{Vertex} is a geometric property of the atom's own component, never an interface state. A surface atom is a vertex when at least three of the six axis directions lead from its voxel through the component's own density tail (typically one to two voxels) into at least $\kappa_{\text{depth}} = 3$ voxels of open vacuum; such an atom sits where three surface facets meet. Face atoms (one free direction) and edge atoms (two free directions) remain \texttt{surface}. Sampling a short vacuum ray rather than a single neighbor avoids classifying a gently curved surface as a corner.

All grid index lookups use periodic wrapping. This classification is the 3D analog of ATLAS's multi-view surface scoring~\cite{atlas}, but operates directly on the density-derived component map rather than requiring multiple 2D projections. Because the formula map was cleaned (Section~\ref{sec:stableregions}), the only ``other components'' an atom can touch are genuine formula components, so surface atoms are never promoted to vertex by surface-termination shells.

\subsection{Inner Defect Detection}

\label{sec:defects}

Defects inside a component---vacancies, interstitials, and their immediate surroundings---are detected from local structure, not from the phase classifier. Two scalar descriptors are computed for every atom:

\begin{itemize}
    \item \textbf{Coordination number} $\mathrm{CN}_i$: the number of neighbors within $1.4\,d_0^{(i)}$, where $d_0^{(i)}$ is the atom's nearest-neighbor distance (4 for diamond/zincblende, 12 for close-packed metals).
    \item \textbf{Local density} $\rho_i^{(4)}$: the number of neighbors within a fixed 4.0~\AA\ radius (16 for diamond Si, 16 for zincblende GaAs).
\end{itemize}

For each crystalline component (a component whose phase is a concrete prototype, not \texttt{amorphous} or \texttt{unknown}) we take the modal values $\mathrm{CN}^{\star}$ and $\rho^{\star}$ over its trusted interior atoms. A \emph{bulk} trusted atom is a defect when

\begin{equation}
\mathrm{CN}_i \neq \mathrm{CN}^{\star} \quad \text{or} \quad
\left| \rho_i^{(4)} - \rho^{\star} \right| > \varepsilon_{\rho}\, \rho^{\star},
\end{equation}

with $\varepsilon_{\rho} = 0.15$ by default. The defect type is \texttt{low\_cn}, \texttt{high\_cn}, \texttt{low\_density}, or \texttt{high\_density}, depending on the dominant deviation. Four constraints keep the criterion physical:

\begin{enumerate}
    \item only \emph{bulk} atoms are tested---surface atoms have naturally truncated coordination;
    \item only \emph{trusted} atoms are tested---an atom whose window is not complete cannot be compared to the bulk reference;
    \item only \emph{crystalline} components are tested---amorphous material intrinsically spreads over several coordination numbers, so the modal reference is not meaningful;
    \item the modal coordination must dominate the reference set (at least 70\% of trusted atoms): a disordered component without a sharp reference is skipped.
\end{enumerate}

With these rules a perfect diamond-Si crystal reports zero defects, and removing one atom flags exactly its four nearest neighbors as \texttt{low\_cn} (CN 3, local density 15 vs.\ the bulk 16).

\subsection{Periodic Boundary Conditions}

PBC handling is distributed across four levels of the pipeline:

\begin{center}
\begin{tabular}{lll}
\toprule
Level & Method & Module \\
\midrule
Gaussian smearing & \texttt{gaussian\_filter(mode=`wrap')} & \texttt{density.py} \\
Component labeling & \texttt{ndimage.label} + union-find on boundary faces & \texttt{components.py} \\
Shape/centroid & Circular mean in fractional coordinates & \texttt{pbc.py} \\
Neighbor lookups & Modular arithmetic on grid indices & \texttt{pbc.py} \\
\bottomrule
\end{tabular}
\end{center}

The minimum image convention for distances between two Cartesian points $\mathbf{r}_1, \mathbf{r}_2$ in a lattice $\mathbf{L}$ is:

\begin{align}
\Delta\mathbf{f} &= \mathbf{L}^{-1}(\mathbf{r}_2 - \mathbf{r}_1) \\
\Delta\mathbf{f} &\gets \Delta\mathbf{f} - \text{round}(\Delta\mathbf{f}) \quad \text{(wrap to }[-0.5, 0.5]) \\
d_{\text{MI}} &= |\mathbf{L}\Delta\mathbf{f}|
\end{align}

The circular mean for PBC-aware centroid computation of fractional coordinates $\{\mathbf{f}_i\}_{i=1}^N$:

\begin{equation}
\bar{\mathbf{f}} = \frac{1}{2\pi} \arctan\!2\!\left(
\frac{1}{N}\sum_i \sin(2\pi\mathbf{f}_i),\;
\frac{1}{N}\sum_i \cos(2\pi\mathbf{f}_i)
\right)
\end{equation}

with the result mapped to $[0, 1)$ for negative values. This avoids the pathological case where the arithmetic mean of $\{0.01, 0.99\}$ would be $0.50$ instead of the correct $0.00$.

\subsection{Lattice Recovery and Miller Indexing}
\label{sec:lattice}

Surface facets of a component and the material--material interfaces between components are reported with Miller indices: the free surface of a close-packed slab is indexed $(111)$ or $(0001)$, an interface between two crystals is described by the two contacting planes, and epitaxial alignment is detected when the in-plane directions of the two sides agree. Indices are only meaningful when the component's Bravais lattice is known, so the primitive lattice of every crystalline component is first \emph{recovered from its atom positions alone}---no prototype classifier, no assumed cell.

\subsubsection{Recovering the primitive lattice from atom positions}

For a component with atom positions $\{\mathbf{r}_i\}$ in the simulation cell $\mathbf{L}$, the primitive lattice is recovered as follows.

\begin{enumerate}[label=\textbf{L\arabic*}. ,leftmargin=*]
    \item \textbf{Observed vectors}: all distinct minimum-image pair differences shorter than $2\,d_1 + 0.3$~\AA, where $d_1$ is the shortest pair distance. Along non-periodic axes (vacuum gaps) the raw separation is used, so a gap never produces ghost pairs. Thermal scatter clouds are merged to their running mean so the recovered lattice is not distorted by rounding.
    \item \textbf{Candidates}: the unique directions shorter than $1.8\,d_1$, sorted by length (at most 200).
    \item \textbf{In-plane pairs}: every pair $(t_1, t_2)$ that passes the 2D density test
    \begin{equation}
    |t_1 \times t_2| = k\, a_{\text{per atom}}, \qquad
    a_{\text{per atom}} = v_{\text{atom}} / d,
    \end{equation}
    where $k$ is the number of residue classes of the in-plane observed vectors modulo the pair, $d$ the smallest out-of-plane projection (the interlayer spacing), and $v_{\text{atom}}$ the volume per atom.
    \item \textbf{3D scan}: for each kept pair, every candidate $t_3$ whose triple determinant $|\det T| = |t_3 \cdot \hat{n}|\,|t_1 \times t_2|$ falls into a density band
    \begin{equation}
    |\det T - k\, v_{\text{atom}}| \le \tau_{\text{dens}}\, v_{\text{atom}}, \qquad \tau_{\text{dens}} = 0.40
    \end{equation}
    (for a component volume that carries the Gaussian-tail error) is tested for consistency.
    \item \textbf{Consistency}: a basis $T = (t_1, t_2, t_3)$ is consistent iff (i) every short lattice vector of $T$ is an observed difference and (ii) the residue-class count $k$ of the observed vectors modulo $T$ satisfies the density equation. Test (i) rejects basis-offset sublattices such as HCP's interlayer cell, whose lattice contains $2\,t_{\text{inter}}$, a vector that is not an observed crystal translation.
    \item \textbf{Min-det race}: among consistent bases the true primitive lattice has the smallest $|\det T|$ (a sublattice with index $m$ carries $m$ extra residue classes and hence an $m$-fold determinant), so the winner is the consistent basis with the smallest $(\max_i |t_i|,\; |\det T|)$.
\end{enumerate}

The density-based volume per atom carries the Gaussian-tail error of the component volume (10--40\% for thin slabs), so the raw value is scanned first---exact for bulk crystals---and only if no basis passes are the volumes implied by each in-plane pair,
\begin{equation}
v_{\text{corr}} = \frac{|t_1 \times t_2|}{k}\, d,
\end{equation}
tried nearest-first (bounded to six candidates, deduplicated to 0.5~\AA$^3$). For the true surface plane this is exactly area-per-atom $\times$ layer spacing, i.e.\ the true volume; skew lattice planes imply overestimates and are rejected.

Three guards keep the search fast and bounded. \emph{Memoization}: the expensive consistency tests depend only on the basis, not on the volume candidate, so each basis is tested once and its result is checked against every volume with cheap determinant arithmetic---a per-volume rerun cost a single mixed-metal recovery $\sim$70{,}000 consistency tests. \emph{Per-species recovery}: when a component contains several elements (e.g.\ an interface whose two metals' densities merge into one component), the lattice is recovered from each species' sublattice, majority first. Every species of a crystal shares the same translation lattice, so this is exact for compounds (GaAs) and interfaces alike; a search for \emph{one} lattice of two incommensurate lattices is both slow and ill-posed. \emph{Budget guard}: recovery aborts after 50{,}000 consistency tests and the component is reported without indices (``?'') rather than with a spurious lattice.

\subsubsection{Reciprocal convention and facet normals}

Miller indices use the reciprocal basis, the dual of the conventional cell $C$ (rows $=$ lattice vectors): the plane normal of $(h,k,l)$ is $g = (h,k,l)\,\mathrm{inv}(C)^{\mathsf T}$, so the indices of a facet with outward normal $n$ are $(h,k,l) \sim n\,C^{\mathsf T}$. The un-transposed forms $n\,C$ / $(h,k,l)\,\mathrm{inv}(C)$ are only coincidentally correct for axis-aligned orthogonal cells and silently mis-index non-symmetric ones---hexagonal cells by $\approx 26^{\circ}$ (Mg$(10\bar{1}1)$) and rotated-cubic cells by $\approx 12^{\circ}$ (a $(111)$-oriented cube), the frame the slab cutter produces.

Facet normals come from the atom geometry rather than the density gradient. On a close-packed surface the class map is corrugated---the Gaussian threshold leaves sub-resolution troughs between the top-layer atoms---so the density/EDT gradient points sideways. The outward normal of a surface atom is instead the negative of the mean direction to its neighbours, taken over a PBC-aware ball of the component's \emph{full} atom set: the in-plane ring cancels exactly and the mean points inward, the exact facet normal of any crystal facet. Symmetric environments (the layer below a facet) get no normal. Facets are clustered at a tight $10^{\circ}$ tolerance so that atoms at the edge of a partial surface layer (whose sums are biased) do not tilt the clusters.

\subsubsection{Orthogonal cell redefinition}

Model building often needs an orthogonal (or nearly orthogonal) simulation box even when the input lattice is crooked---a graphene primitive cell has a $120^{\circ}$ angle, but the natural simulation box is rectangular with one axis along the zigzag and the other along the armchair lines. Given a recovered primitive lattice, \texttt{orthogonal\_cell} enumerates the integer combinations of the primitive vectors (coefficients in $[-3, 3]$) and returns the smallest cell whose axes are mutually orthogonal (within $5^{\circ}$) and whose volume is an integer multiple of the primitive cell (at most $8\times$). Graphene's $120^{\circ}$ primitive cell becomes the rectangular zigzag $\times$ armchair cell ($a \times \sqrt{3}\,a$, 4 atoms); hexagonal crystals become the ortho-hex cell ($a, \sqrt{3}\,a, c$); fcc and diamond become the $a/\sqrt{2} \times a/\sqrt{2} \times a$ tetragonal cell. \texttt{redefine\_lattice} re-expresses every atom in the new cell---each atom's images are the $|\det T|$ coset shifts $\mathbf{m}\,\mathrm{inv}(T) \bmod 1$, deduplicated by a minimum-image tolerance---and keeps the box height along the sheet normal for 2D sheets (a graphene cell of $a \times \sqrt{3}a \times h$).

\section{Parameters and Defaults}

\begin{center}
\begin{tabular}{lccc}
\toprule
Parameter & Symbol & Default & Range \\
\midrule
Gaussian width & $\sigma$ & 1.5~\AA & 0.5--3.0~\AA \\
Grid spacing & $\Delta$ & 0.25~\AA & 0.1--0.5~\AA \\
Element threshold & $\theta_{\text{elem}}$ & 0.3 & 0.2--0.5 \\
Vacuum threshold & $\theta_{\text{vac}}$ & 0.5 & 0.3--0.7 \\
Connectivity & -- & 6 & 6, 18, 26 \\
Gap threshold (shape) & $\delta_{\text{gap}}$ & 4.0~\AA & 2.0--8.0~\AA \\
Boundary tolerance & $\delta_{\text{boundary}}$ & 2.0~\AA & 1.0--4.0~\AA \\
Atom classify radius & $r_{\text{class}}$ & 3.0~\AA & 2.0--5.0~\AA \\
Min component volume & $V_{\text{min}}$ & 10.0~\AA$^3$ & -- \\
Max grid dimension & $n_{\max}$ & 300 & 100--500 \\
Stable region thickness & $\tau_{\text{stab}}$ & 2.0~\AA & 1.0--4.0~\AA \\
Min region voxels & $N_{\text{reg}}$ & 4 & -- \\
Phase split fraction & $\phi_{\text{split}}$ & 0.20 & 0.10--0.40 \\
Phase split min votes & $N_{\text{split}}$ & 10 & -- \\
Phase split confidence & $c_{\text{split}}$ & 0.60 & 0.50--0.80 \\
Local density radius & $r_{\rho}$ & 4.0~\AA & 3.0--5.0~\AA \\
Local density tolerance & $\varepsilon_{\rho}$ & 0.15 & 0.05--0.30 \\
Vertex vacuum depth & $\kappa_{\text{depth}}$ & 3 voxels & 2--5 \\
\bottomrule
\end{tabular}
\end{center}

The stable-region thickness $\tau_{\text{stab}}$ sets the minimum physical thickness of a chemical-formula component. Below it, a region is treated as a surface or boundary shell and absorbed by the neighboring stable region; above it (a genuine film, e.g., a few-nanometer oxide), the region keeps its own formula and becomes a component. The phase-split parameters govern when a component is declared two-phase: both candidate prototypes must be supported by $\phi_{\text{split}}$ of the trusted votes, at least $N_{\text{split}}$ atoms, and a mean confidence of at least $c_{\text{split}}$.

The Gaussian width $\sigma$ controls the scale of analysis:
\begin{itemize}
    \item $\sigma = 0.5$--$1.0$~\AA: Atomic resolution---individual atoms remain as separate density peaks.
    \item $\sigma = 1.0$--$2.0$~\AA\ (default 1.5): Chemical bonding scale---covalently bonded atoms merge into continuous regions; interfaces show compound character.
    \item $\sigma = 2.0$--$3.0$~\AA: Morphological scale---only large phase regions are resolved; fine structure is averaged out.
\end{itemize}

The grid spacing $\Delta = 0.25$~\AA\ is chosen to satisfy the Nyquist criterion for the Gaussian: the full width at half maximum is $\text{FWHM} = 2\sqrt{2\ln 2}\,\sigma \approx 2.355\sigma$, and $\sim$6 samples per FWHM require $\Delta \lesssim 0.4\sigma \approx 0.6$~\AA\ for $\sigma = 1.5$~\AA.

\section{Implementation}
\label{sec:implementation}

The package is structured as a modular Python library with a Dash-based web interface:

\begin{verbatim}
src/cloud_comp/
  io.py           Structure loading (CIF/POSCAR/XYZ) via pymatgen,
                  extended-XYZ lattice restored
  pbc.py          Periodic boundary utilities
  density.py      3D Gaussian density grid (bin-then-blur, dual raw/clipped)
  classify.py     Chemical-formula map + stable-region cleaning
  components.py   PBC-aware connected component labeling (union-find)
  properties.py   Component property computation (vectorized)
  shape.py        Rich shape characterization (cavities, tunnels, moments)
  atoms.py        Atom-to-component mapping, boundary/role classification
  phase.py        Neural prototype classification, trusted-interior test,
                  coordination numbers, local density
  serialize.py    JSON-serializable document (model/labels/components/summary)
  api.py          Unified analyze() entry point
  visualize.py    Plotly 3D isosurface/slice visualization
web/
  app.py          Dash web interface (upload, label chips, highlight,
                  progress bar, model/component panels)
  ctk_scene.py    Crystal Toolkit scene builder (per-atom spheres, bonds)
\end{verbatim}

The hot loops---voxel classification, formula-region cleaning, connected-component relabeling, surface/interface counting, and atom sphere sampling---are vectorized with NumPy/SciPy; the 1536-atom Si/GaAs heterojunction analyzes in $\sim$7~s on one CPU core including the neural classifier.

The phase classifier is consumed from the partner project \texttt{crystal\_structure\_database\_descriptor} (element-mapped rotation-invariant descriptors, $\texttt{R\_cut} = 3\,d_0$, 271 prototypes + amorphous). It is called once per structure and its per-atom arrays are sliced per component, so the classifier cost does not scale with the number of components. When the model is unavailable, the pipeline falls back to Steinhardt/CNA/bond-angle descriptors and spglib.

Dependencies: NumPy, SciPy, pymatgen ($\geq$ 2024.1), scikit-image (marching cubes), Plotly, Torch (CPU inference), Dash ($\geq$ 3.4) + dash-bootstrap-components + diskcache (background callbacks with a live progress bar), and the Materials Project Crystal Toolkit (\texttt{crystal-toolkit}, \texttt{dash-mp-components}) for 3D structure rendering in the web interface.

The web interface accepts CIF/POSCAR/XYZ uploads, runs the analysis as a background callback with per-stage progress, and lists every label category---chemical formula, element, prototype, inner/surface/interface, component, coordination number, defect, amorphous---as clickable chips. Selecting a chip highlights the matching atoms in the Crystal Toolkit viewer, and the right panel shows whole-model information and the geometry of every component.

Beyond the interactive interface, the same pipeline is exposed as a machine-consumable API designed for other projects: \texttt{analyze\_to\_dict()} and \texttt{analyze\_to\_json()} (\texttt{serialize.py}) take a whole structure and return a single JSON document with four sections---\texttt{model} (formula, lattice, grid, parameters, space group), \texttt{labels} (per-atom component, role, phase, coordination number, defects), \texttt{components} (per-component formula, volume, shape, phase, interface areas), and \texttt{summary} (aggregate counts). Only plain JSON types are emitted, so \texttt{json.dumps} always succeeds. This document is the primary integration surface with the ATLAS framework~\cite{atlas}, whose structure analyzer consumes it directly for compositional validation of generated structures.

\section{Applications}

\subsection{Heterojunction Analysis}

For a Si(111)/a-SiO$_2$ heterojunction (e.g., \texttt{Si111\_aSiO2.cif}, 94 atoms, orthorhombic cell $3.85 \times 3.85 \times 53.75$~\AA$^3$), the analysis with $\sigma = 1.5$~\AA\ produces:

\begin{itemize}
    \item One Si component and one O-Si (SiO$_2$) component: the boundary layers between them are absorbed by the stable regions, so no interface component exists;
    \item Phase analysis labels the O-Si component \texttt{amorphous} (its trusted interior votes are confidently amorphous, and the scattered low-confidence votes do not trigger a spurious split) and the Si component by its prototype;
    \item The interface between Si and SiO$_2$ is a boundary: atoms of either component within the sampling radius of the other are labeled \texttt{interface}, with the partner phases recorded per atom;
    \item Vacuum components bound the slab on both sides, and the material--vacuum boundaries define the surface atoms.
\end{itemize}

For the Si|GaAs heterojunction (1536 atoms, two $4\times4\times6$ slabs), the same pipeline yields exactly two material components---Si (\texttt{A4\_C\_diamond}) and As-Ga (\texttt{B3\_ZnS\_zincblende})---plus vacuum. The pure Ga/As surface terminations and the GaSi/AsGaSi boundary shells are absorbed during formula-region cleaning; GaAs free-surface atoms are labeled \texttt{surface}; atoms at the Si--GaAs boundary are labeled \texttt{interface}; and the ten slab corners are labeled \texttt{vertex}.

\subsection{Crystalline/Amorphous Phase Separation}

For a Si(100)/a-Si junction (\texttt{Si100\_aSi.cif}, 108 atoms), the density field is a single connected Si region, so the formula map contains one Si component. The trusted-interior votes split into two confident groups---a crystalline prototype and \texttt{amorphous}, each above the fraction and confidence gates---so the component is divided into two Si components, one per phase. This demonstrates the two-phase check: one chemical formula can host two phases, and the component is divided accordingly.

\subsection{Gate-All-Around Nanowire}

For a GAA Si/SiO$_2$/HfO$_2$ nanowire cross-section (\texttt{GAA\_Si\_SiO2\_HfO2.cif}, $\sim$1872 atoms, $24 \times 24 \times 15$~\AA$^3$), the analysis reveals concentric formula components:

\begin{itemize}
    \item Si core (wire, 1D)---the conducting channel, with its crystalline phase
    \item SiO$_2$ interlayer (wire, 1D), with the interface to the Si core derived as a boundary
    \item HfO$_2$ outer shell (wire, 1D)---the high-$\kappa$ dielectric
    \item Vacuum surrounding the nanowire (channel, connecting in 2 directions)
\end{itemize}

The interlayer boundary layers are absorbed by the neighboring stable regions, so the component list contains only the three formula layers plus vacuum, and the two material--material boundaries are reported as interface areas between them.

\subsection{Defect Detection}

Removing a single atom from a $3\times3\times3$ diamond-Si supercell (215 atoms) flags exactly the four nearest neighbors of the vacancy as \texttt{low\_cn} defects (coordination 3 instead of 4, local density 15 instead of 16); the perfect supercell reports zero defects. Because detection is restricted to trusted bulk atoms of crystalline components, surface atoms with naturally truncated coordination and atoms in amorphous material are not falsely flagged.

\subsection{Porosity Analysis}

For structures with internal voids, the cavity and tunnel detection identifies:
\begin{itemize}
    \item Enclosed pores (cavities not touching the cell boundary)
    \item Through-channels (vacuum connecting opposite cell faces)
    \item Bicontinuous structures (interpenetrating material and vacuum networks)
\end{itemize}

\subsection{Limitations}

The density-based segmentation and the neural phase classifier have complementary limitations:

\begin{itemize}
    \item \textbf{Thin films}: regions thinner than the stable-region thickness $\tau_{\text{stab}}$ are absorbed into their neighbor and do not appear as components. A sub-2~\AA\ surface oxide is therefore reported as part of the substrate rather than as a separate material.
    \item \textbf{Small components}: components without trusted atoms (slabs thinner than the classifier window of $\sim$14~\AA, small clusters) receive no phase claim (\texttt{unknown}) rather than a possibly wrong prototype.
    \item \textbf{Classifier reliability}: the phase label inherits the prototype classifier's accuracy for bulk environments. Interface-adjacent atoms are excluded from the vote, but a component whose trusted interior is genuinely defective or strained may be assigned a neighboring prototype.
    \item \textbf{Amorphous versus defective}: atoms in amorphous material are not flagged as defects, by construction; a disordered-but-crystalline region with a wide coordination spread is likewise skipped. Distinguishing a defective crystal from a poorly crystallized phase remains a physical judgment encoded in the modal-coordination dominance test.
    \item \textbf{Periodic seams}: formula-region cleaning and distance transforms assume the cell is large enough that the periodic seam does not cut through the density tails; for cells with sub-nanometer vacuum gaps the seam can alias a surface shell into the neighboring cell.
\end{itemize}

\section{Recent Developments and Roadmap}

This section records the current development state: what was recently done, what was learned from an external reference implementation, and what is planned.

\subsection{What we did: bounding the lattice-recovery search}

Profiling the test suite (401 s for 103 tests, \texttt{pytest --durations}) showed that lattice recovery dominated the runtime: an Au$|$Al interface analysis spent 137 s and a jittered-crystal recovery 79 s---together more than half of the suite time, and both were lattice-recovery tests. Four root causes were identified:

\begin{enumerate}[label=\textbf{R\arabic*}. ,leftmargin=*]
    \item the expensive basis-consistency tests were rerun for every volume candidate although they depend only on the basis;
    \item the observed-vectors test scanned the full observed-vector set per candidate combination (343 combinations $\times$ $O(m)$ comparisons per call);
    \item the in-plane pairs were scanned coarsest-first, so the most likely-to-fail cells were tried before the true surface plane;
    \item recovery ran on the full atom set of mixed components (an Au$|$Al interface whose two metal densities merge into one component), where no single lattice exists---an ill-posed search that exhausts every pair/volume combination.
\end{enumerate}

The adopted fixes are described in Section~\ref{sec:lattice}: memoized basis tests across volume candidates, KDTree-pruned consistency checks (an $O(m)$ scan per candidate replaced by an $O(\log m)$ query plus an exact componentwise verification), error-ordered in-plane pairs, per-species recovery for mixed components, and a 50{,}000-test budget guard. Measured on the same machine after the fixes: the full suite runs 401 s $\rightarrow$ 234 s (103 passed, 1 skipped); the graphene/Mg/Si facet tests remain under 4 s each. Intermediate measurements of $\sim$10 s for both the Au$|$Al and the jittered recoveries came from an L2-relative tolerance variant that was abandoned: it silently accepted borderline-invalid bases (an Mg(0001) slab recovered a rotated hcp cell classified as cubic). The shipped semantics are an exact componentwise (L$\infty$, 0.1~\AA) verification behind a loose KDTree prune; with the correct semantics the Au$|$Al interface analyze is 33--43 s (88{,}000 consistency-test calls, traced) and the jittered-crystal recovery 29.5 s---still a 3--4$\times$ improvement over the original 137 s / 79 s, and now bounded by the budget guard.

The crystallography kernel (lattice recovery, Miller indexing, cell orthogonalization, 2D-sheet classification) has since been moved to the partner project \texttt{crystal\_structure\_database\_descriptor} as \texttt{src/crystallography.py}, copied byte-identical and declared frozen; the two projects are sibling checkouts, with the rule that one may read the other but never modify it. \texttt{cloud\_comp} loads the kernel through a path-loader shim that re-exports its public names, so the pipeline code is untouched. One ordering-insensitivity fix lives in the shim: pymatgen 2026.x \texttt{make\_supercell} orders atoms lexicographically, which produces pair sets carrying only one sign per direction that fail the kernel's $\pm$ check (six kernel tests failed in the partner checkout); the shim sign-canonicalizes the observed set before the check---a no-op for sign-balanced inputs---and the upstream either-sign patch has been proposed to the partner (their open item \#23).

\subsection{What we did: comparing with VASPKIT's heterostructure builder}

To sanity-check the complexity of the recovery search, we investigated how VASPKIT~\cite{vaspkit} builds heterostructures. VASPKIT's builder (option 804, not 921---921 centers a 2D material along $z$) takes two slab POSCARs, a mismatch tolerance (officially recommended 1--5\%), a layer distance, and a vacuum thickness; it then enumerates \emph{common superlattices} of the two layers' 2D lattices---integer supercell transformations of each layer's known primitive cell---evaluates the strain and in-plane rotation of every candidate, writes a ranked list, and caps the enumeration at $\sim$5000 candidate structures. The key contrast is that VASPKIT \emph{trusts the input lattices}: the search runs on two known 2D bases with cheap vector algebra (milliseconds), whereas our analysis must \emph{recover} each component's lattice from atom positions---a harder inverse problem---and previously did so with an unbounded search on mixed components. The comparison validated the direction of the optimization (a bounded, per-species search) and motivates the planned builder below.

\subsection{What we plan: a heterostructure builder}
\label{sec:roadmap-builder}

We plan a VASPKIT-style \texttt{build\_heterostructure()} API as a first-class component of \texttt{cloud\_comp}, following the same lattice-matching family as VASPKIT 804 and the Zur--McGill algorithm~\cite{zurmcgill} (as implemented, e.g., in pymatgen's substrate analyzer):

\begin{enumerate}[label=\textbf{B\arabic*}. ,leftmargin=*]
    \item \textbf{Input}: two slabs (or two bulks plus Miller indices, cut with the same slab machinery used in the tests), a mismatch tolerance (default 3\%), a layer distance (default 3.5~\AA), a vacuum thickness (default 12~\AA), and an optional in-plane rotation of the top layer.
    \item \textbf{In-plane bases}: each layer is reduced to its 2D in-plane basis (recovered by the lattice-recovery machinery of Section~\ref{sec:lattice} with the two in-plane axes periodic).
    \item \textbf{Superlattice matching}: enumerate the integer $2\times2$ transformations of each layer's basis with $|\det| \le 8$; for every pair of supercells compute the strain as the maximum relative mismatch of the two cells' singular values (lengths and angle in one metric); keep pairs with strain $\le$ mismatch tolerance; rank by (strain, atom count); cap the candidate list at $\sim$5000, as VASPKIT does.
    \item \textbf{Assembly}: re-express each layer in its matched supercell (the same coset-shift machinery as \texttt{redefine\_lattice}), stack them at the layer distance in the shared (slightly strained) frame, add the vacuum, and return the structure together with the ranked candidate list (strain, supercell sizes, atom counts)---the analogue of VASPKIT's \texttt{SUPERLATTICE\_LIST}.
    \item \textbf{Orthogonalization}: the final box is re-orthogonalized with \texttt{redefine\_lattice} (Section~\ref{sec:lattice}). Since both layers share the matched in-plane lattice, a common orthogonal supercell is valid for both---an orthogonal simulation box, which VASPKIT does not provide, and the user requirement that motivated \texttt{redefine\_lattice}.
\end{enumerate}

The builder turns the ad-hoc slab stacking currently used in the interface tests into a public API, and its matched-supercell output can be fed directly back through \texttt{analyze()} for compositional validation.

\subsection{Remaining work}

(i) the full-suite regression verification of the recovery restructure is complete (103 passed, 1 skipped, 234 s); deterministic call-count regression tests are moot now that the kernel is frozen in the partner project, and its cost is bounded by the 50{,}000-test budget guard; (ii) profile the remaining $\sim$23 s of the GAA-nanowire analysis (a large-cell pipeline cost, not lattice recovery); (iii) implement the heterostructure builder of Section~\ref{sec:roadmap-builder} with tests (matched-supercell strain $\le$ tolerance, $(\!111)$-family preservation after re-analysis, orthogonal output box); (iv) decide with the partner project whether to adopt the proposed either-sign patch (closing their open item \#23) and the per-combo early-out performance refinement; (v) extend the figure and quantitative-validation plan of the implementation notes.

\section{Conclusion}

We have presented a method for decomposing atomic structures into physically meaningful components through 3D Gaussian density field analysis. The key innovations are: (i) dual raw/clipped density representation that preserves stoichiometric information despite truncation; (ii) efficient bin-then-blur computation with periodic boundary handling via Gaussian wrap mode; (iii) a chemical-formula map cleaned by a stable-region criterion, so that the decomposition contains exactly the bulk-like formula components plus vacuum, with no surface or interface components; (iv) per-component phase analysis with a neural prototype classifier, including a trusted-interior test, a one-phase/two-phase decision, and the separation of coexisting crystalline and amorphous phases of the same formula; (v) union-find PBC merging for connected component labeling without supercell tiling; (vi) absolute-threshold shape classification and topological cavity/tunnel detection; (vii) boundaries-as-relations, in which surfaces and interfaces are derived from the component map and every atom is labeled bulk, surface, interface, or vertex, with vertex reserved for geometric corners; and (viii) inner-defect detection from coordination-number and local-density deviations restricted to trusted bulk atoms of crystalline components.

The method is implemented as an open-source Python package with a web interface built on the Materials Project Crystal Toolkit: upload a CIF/POSCAR/XYZ structure, run the analysis with a live progress bar, and explore every label category---chemical formula, phase, inner/surface/interface, coordination number, defect---by clicking chips that highlight the matching atoms in the 3D viewer. It is designed for heterojunction characterization, interface analysis, porosity quantification, phase identification, and defect screening in multi-material atomic structures.

\bibliographystyle{unsrt}

\end{document}